\def\PRX{{ Phys. Rev. X\ }\/}
\def\PRL{{ Phys. Rev.Lett.\ }\/}
\def\PRB{{ Phys. Rev. B\ }\/}

\def\be{\begin {equation}}
\def\ee{\end {equation}}
\def\ber{\begin {eqnarray}}
\def\eer{\end {eqnarray}}
\def\bers{\begin {eqnarray*}}
\def\eers{\end {eqnarray*}}

\makeatletter

\newcommand{\Rmnum}[1]{\expandafter\@slowromancap\romannumeral #1@}
\makeatother

\makeatletter
\newcommand*\env@matrix[1][*\c@MaxMatrixCols c]{%
  \hskip -\arraycolsep
  \let\@ifnextchar\new@ifnextchar
  \array{#1}}
\makeatother

\documentclass[aps,prb,twocolumn,showpacs,groupedaddressamsmath,amssymb]{revtex4-1}

\usepackage{amsmath}
\usepackage{float}
\usepackage{color}
\usepackage[normalem]{ulem}
\usepackage{amsfonts}
\usepackage{amssymb}
\usepackage{graphicx}
\begin {document}

\title{Emergence of Topological insulator and Nodal line semi-metal states in XX$^{'}$Bi (X=Na, K, Rb, Cs; X$^{'}$=Ca, Sr)}

\author{ Chiranjit Mondal,$^{1}$ C. K. Barman,$^{2}$ Sourabh Kumar,$^{1}$  Aftab Alam$^{2}$ and Biswarup Pathak$^{1}$}
\email{{\it biswarup@iiti.ac.in}, aftab@.iitb.ac.in}
\affiliation{$^{1}$Discipline of Metallurgy Engineering and Materials Science, IIT Indore, Indore}
\affiliation{$^{2}$Department of Physics, Indian Institute of Technology Bombay, Powai, Mumbai 400076, India}

\pacs{73.20.-r,71.70.Ej,71.15.Mb} 

\begin{abstract}
In this letter, we predict the emergence of non-trivial band topology in the family of XX$^{'}$Bi compounds having $P\overline{6}2m$ (\# 189) space group. Using first principles calculations within hybrid functional framework, we demonstrate that NaSrBi and NaCaBi are strong topological insulator under controlled band engineering. Here, we propose three different ways to engineer the band topology to get a non-trivial order: (i) hydrostatic pressure, (ii) biaxial strain (due to epitaxial mismatch), and (iii) doping. Non-triviality is  confirmed by investigating bulk band inversion, topological Z$_2$ invariant, surface dispersion and spin texture. Interestingly, some of these compounds also show a three dimensional topological nodal line semi-metal (NLS) state in the absence of spin orbit coupling (SOC). In these NLS phases, the closed loop of band degeneracy in the Brillouin zone lie close to the Fermi level. Moreover, a drumhead like flat surface state is observed on projecting the bulk state on the [001] surface. The inclusion of SOC opens up a small band gap making them behave like a topological insulator.
\end{abstract} 
\date{\today}
\maketitle


Symmetry protected nontrivial band topology has become an area of paramount research interest for unravelling novel dimensions in condensed matter physics.\cite{in_1, in_2} The time reversal invariant topological insulator (TI) has stimulated intense interests due to their intriguing properties, such as gapless boundary states, unconventional spin texture and so on.\cite{in_3, in_4, in_5} The recent years have witnessed a series of theoretical developments which have enabled us to classify the $Z_2-$type non-magnetic band insulators. For example, the $Z_2-$even (ordinary) and $Z_2-$odd (topological) states are separated by a topological phase transition, where the bulk gap diminishes during the adiabatic deformation between these two states.\cite{in_6} In two-dimensional (2D) systems, $Z_2-$odd class can be distinguished by the odd number of Kramer's pairs of counter propagating helical edge states, whereas in three-dimensional (3D) systems, it can be characterized by the odd number of Fermi loops of the surface band that encloses certain high symmetry points in the Brillouin zones (BZ).\cite{in_3} Soon after the experimental realization of quantum spin hall effect in 2D HgTe quantum wall,\cite{in_4} a number of 2D and 3D TI systems have been theoretically predicted and experimentally verified.\cite{in_7, in_8, in_9, in_10} In fact, the search for new TI has been extended to zintl compounds,\cite{in_11, in_12} antiperovskites,\cite{in_13} and heavy fermion f-electron Kondo type of systems.\cite{in_14}

With the conceptual development in the topological field, research on topological material has been extended from insulators to semimetals and metals.\cite{in_15, in_16, in_17} In topological semimetals, symmetry protected band crossing or accidental band touching leads to a nontrivial band topology in 3D momentum space. The topological properties of such semimetals mainly depend on the degeneracy of the bands at the crossing/touching point. A zero dimensional band crossing with two and four fold band degeneracy defines the Dirac\cite{in_15} and Weyl semimetal,\cite{in_16} respectively, which are quasi-particle counterparts of Dirac and Weyl fermions in high energy physics. Low energy Dirac fermions in condensed matter are essentially protected by time reversal symmetry (TRS), inversion symmetry (IS) and certain crystal symmetry. Quasi-particle Weyl fermion state can be realized by breaking either space inversion or time reversal of crystal.\cite{in_16} On the contrary, in quantum field theory, Dirac and Weyl fermions are strictly restricted by Lorentz invariance. However, in case of nodal line semimetal (NLS), the conduction and valance band touches along a line to form a one dimensional close loop.\cite{in_17} The characteristic feature of Dirac semimetal (DS) is a point like Fermi surface (FS) at the crossing point,\cite{in_15} whereas it is 1D circle for NLS.\cite{in_17} But for Weyl semimetal (WS), FS forms an arc like surface, instead of closed. \cite{in_16} Due to the nontrivial FSs, all the topological semimetals show some exotic phenomena, such as quantum magneto-resistance,\cite{in_18} chiral anomaly\cite{in_19} etc.

NLS are the precursor states for other topological phases. In general, spinful nodal lines are not robust in the presence of a mass term in Hamiltonian,\cite{in_20} which can be explained by simple co-dimensional analysis. Thus, inclusion of spin orbit coupling (SOC) can convert the NLS state to DS, WS or TI by opening up a gap around the nodal loop. However, in the presence of an extra crystalline symmetry, nodal line can be robust.\cite{in_20} Owing to the unique properties such as torus-shaped Fermi surface, relatively higher density of states and interaction induced instability of the FS, NLS can provide a unique playground for the quasi-particle correlations and unusual transport studies.\cite{in_17}

\begin{figure}[t]
\centering
\includegraphics[scale=0.085]{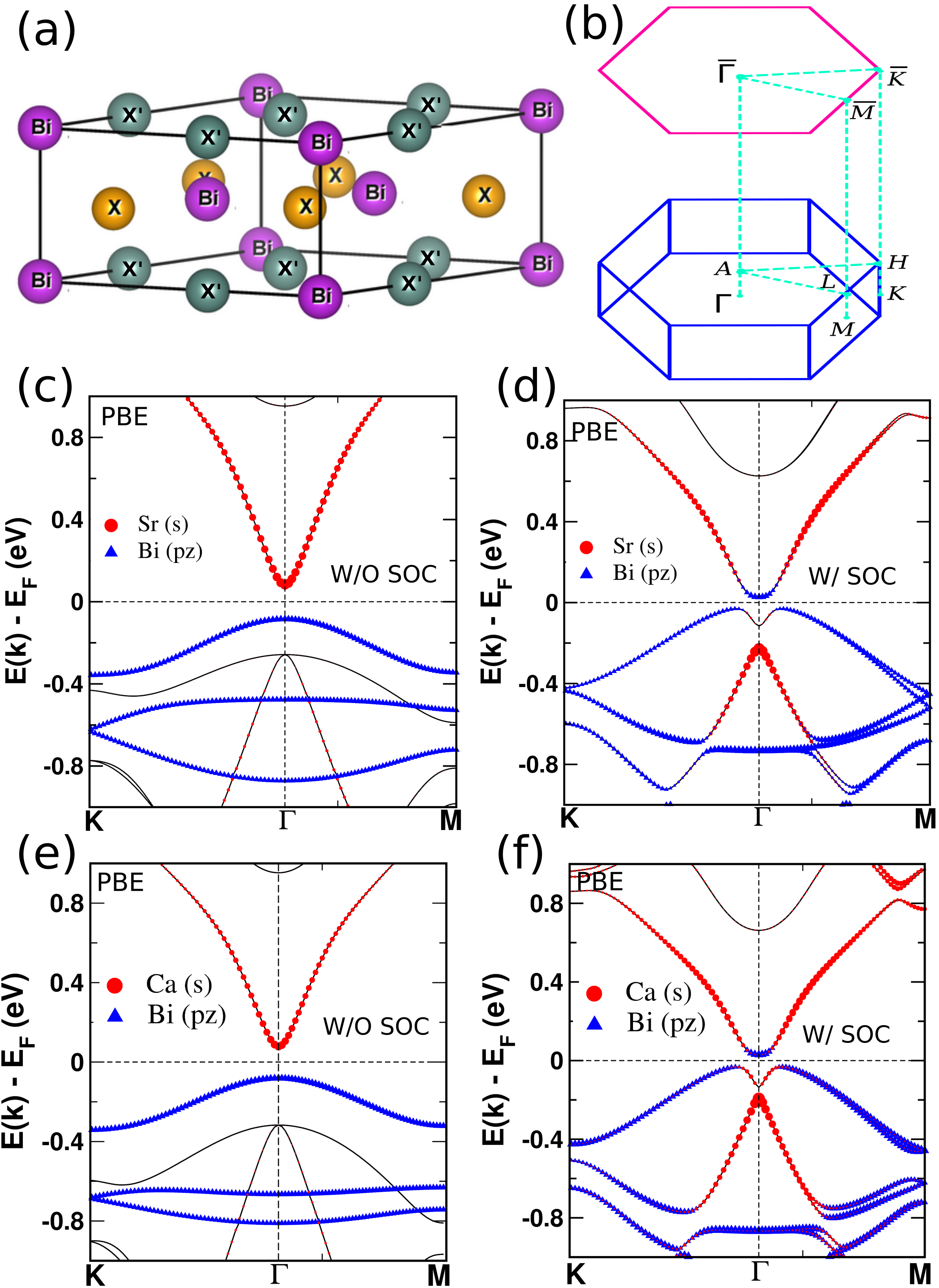}\ 
\caption {(a) Hexagonal crystal structure of XX$^{'}$Bi, (b) Corresponding Brillouin zone (BZ) and two-dimensional BZ projected onto (001) surface. Electronic structures of NaSrBi (c-d) and NaCaBi (e-f) with (W/) and without (W/O) SOC. Red and blue symbols in (c-f) indicate the orbital contributions of Sr/Ca s orbital and Bi $p_z$ orbital. The sizes of the symbols are proportional to the weightage of the orbitals. }
\label{fig1}
\end{figure}

{\par}In this article, we explore the possibility of controlling topological order in a series of ternary compounds XX$^{'}$Bi (X = Na, K, Rb, Cs and X$^{'}$ = Ca, Sr). The  XX$^{'}$Bi compounds have a non-centrosymmetric hexagonal structure with $P\overline{6}2m$ (\# 189) space group as shown in Fig.~\ref{fig1}(a). The theoretically relaxed lattice parameters and formation energy of these systems are provided in supplementary material (SM).\cite{supl} These compounds show interesting topological properties (including NLS state) which can be tuned under various external factors. Recently, NaBaBi has been theoretically studied and predicted to be a topological insulator under hydrostatic pressure.\cite{in_21} 

{\it Computational Details :}
All the calculations were carried out using projector augmented wave \cite{cm_3, cm_4} formalism based on Density Functional Theory (DFT) as implemented in the Vienna Ab Initio Simulation Package (VASP). \cite{cm_1} The Perdew-Burke-Ernzerhof (PBE) \cite{cm_2} type functional with generalized-gradient approximation (GGA) \cite{cm_5} was employed to describe the exchange and correlation effects. All the structures are fully relaxed until the Hellmann$-$Feynman forces on each atom are less than 0.01 eV/{\AA} and the total energy converge up to 10$^{-6}$ eV. An energy cutoff of 500 eV is used to truncate the plane-wave basis sets for the representation of Kohn$-$Sham wave functions. The BZ is  integrated over $7\times7\times11$ gamma centered k-mesh in all the electronic calculations. Hybrid functional (HSE06)\cite{cm_6, cm_7} level of calculations is further carried out to verify the accuracy of PBE-results for electronic structure calculations. Tight-binding (TB) Hamiltonians are constructed using wannier90 package\cite{cm_9} based on the maximally localized Wannier functions \cite{cm_10} (MLWFs). The topological properties including surface spectrum and Fermi surface were analyzed based on the iterative Green's function method.\cite{cm_11} Our formations energy calculations show that the studied systems are energetically stable (Table S1).\cite{supl}


{\it Topological insulator :}

{\par}Topological insulating state can be predicted in a material if it shows band inversion driven by spin-orbit coupling.\cite{in_5} Since the strength of spin-orbit coupling increases with heavy elements, we have systematically studied X-X$^{'}$ elements from group IA-IIA of the periodic table. The electronic structures of all these compounds have been performed using the GGA level of theory. Here, we have mainly discussed on NaSrBi and NaCaBi compounds. Our findings related to all other compounds given in SM.\cite{supl} Fig.~\ref{fig1}(c-f) presents the electronic structers of NaSrBi and NaCaBi compounds. In the absence of SOC, the conduction band minima (CBM) and valence band maxima (VBM) at $\Gamma$ point are dominated by Sr/Ca $s$ and Bi $p_{z}$ orbital as shown in Fig.~\ref{fig1}(c,e). However, inclusion of SOC results in an inverted band order between Sr/Ca $s$ and Bi $p_z$ orbitals at $\Gamma$ point with a direct band gap of $\sim$80(100) meV at $\Gamma$ point for NaSrBi(NaCaBi) as shown in Fig.~\ref{fig1}(d,f) which clearly indicates the non-trivial band topology in these systems. 

{\par} The non-trivial band topology suggests an interesting surface state.\cite{in_2} Henceforth, we have studied surface electronic structure of NaSrBi and NaCaBi compounds. Initially, we constructed the slab Hamiltonian from maximally localized wannier functions (MLWF) for Na $s$, Ca/Sr $s$ and Bi $p$ orbitals. Then we have projected the band structure onto the (001) surface by using the iterative Green's function method as implemented in Wannier-tool.\cite{cm_11} The surface spectra of the slab with a thickness of 200 unit cells are shown in SM (Fig. S1).\cite{supl}

\begin{figure}[h]
\centering
\includegraphics[width=\linewidth]{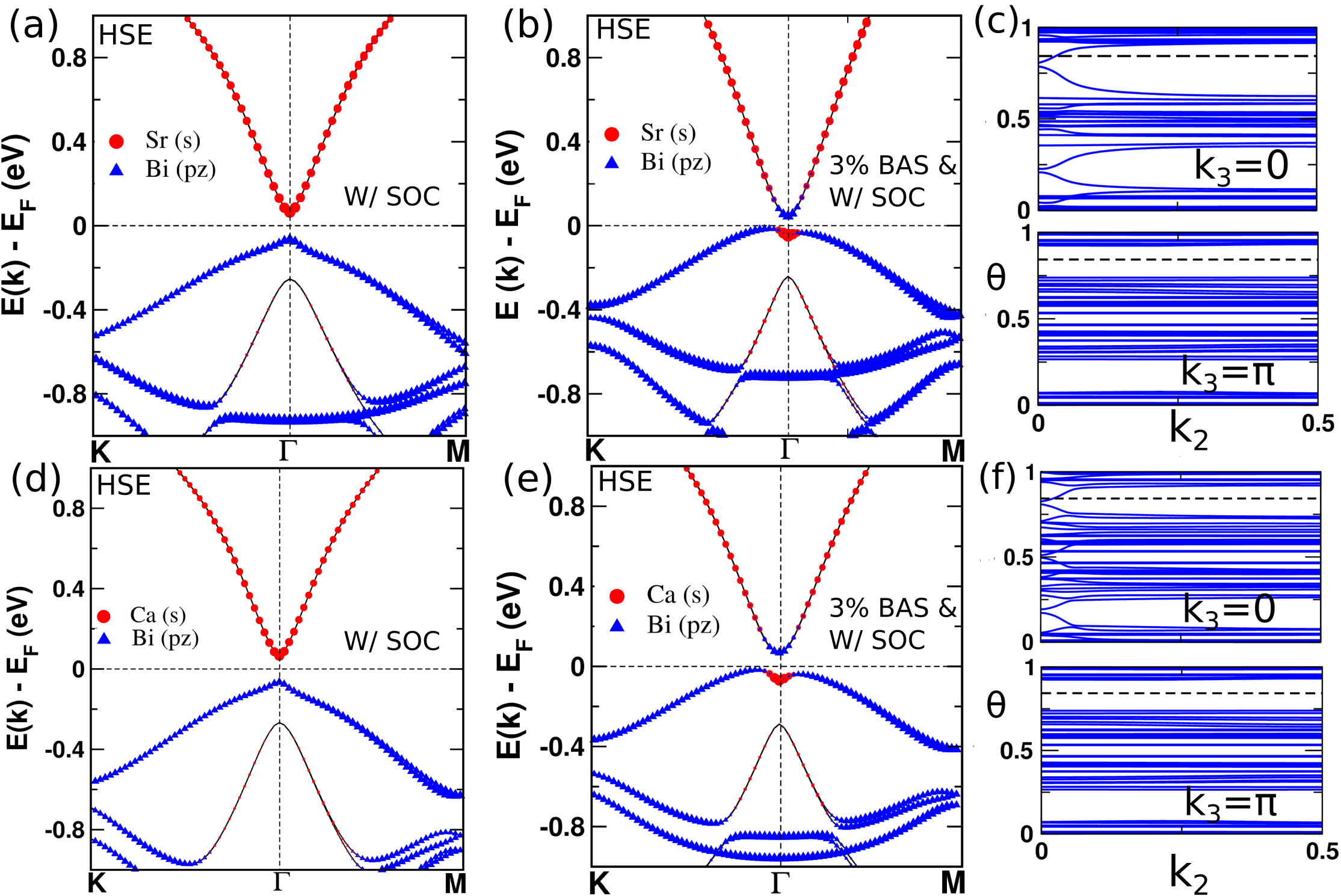}
\caption { Band structures (using HSE06) of (a) pure and (b) 3\% biaxial strained NaSrBi. (c) evolution of wannier charge center along k$_2$ for NaSrBi. (d-f) Respective plots for NaCaBi}
\label{fig2}
\end{figure}

{\par} Since the GGA method underestimates the band gap and overestimates the band inversion, we have used the hybrid functional HSE06 \cite{cm_6, cm_7} to confirm the predicted non-trivial topology. It turns out that the band inversion between Sr/Ca $s$ orbital and Bi $p_{z}$ orbital disappears at HSE06 level and both of these materials show trivial band order as shown in Fig.~\ref{fig2}(a,d). In order to check the evolution of non-trivial band order, we have applied external effects such as pressure, strain and doping. We find that NaSrBi (NaCaBi) system shows a topological insulating behaviour under strain (both hydrostatic as well as biaxial strain induced by epitaxial mismatch). Topological non-trivial properties also emerge in these materials if we partially/fully substitute Na by  K, Rb, and Cs in these compounds.


\begin{figure}[b]
\centering
\includegraphics[width=\linewidth]{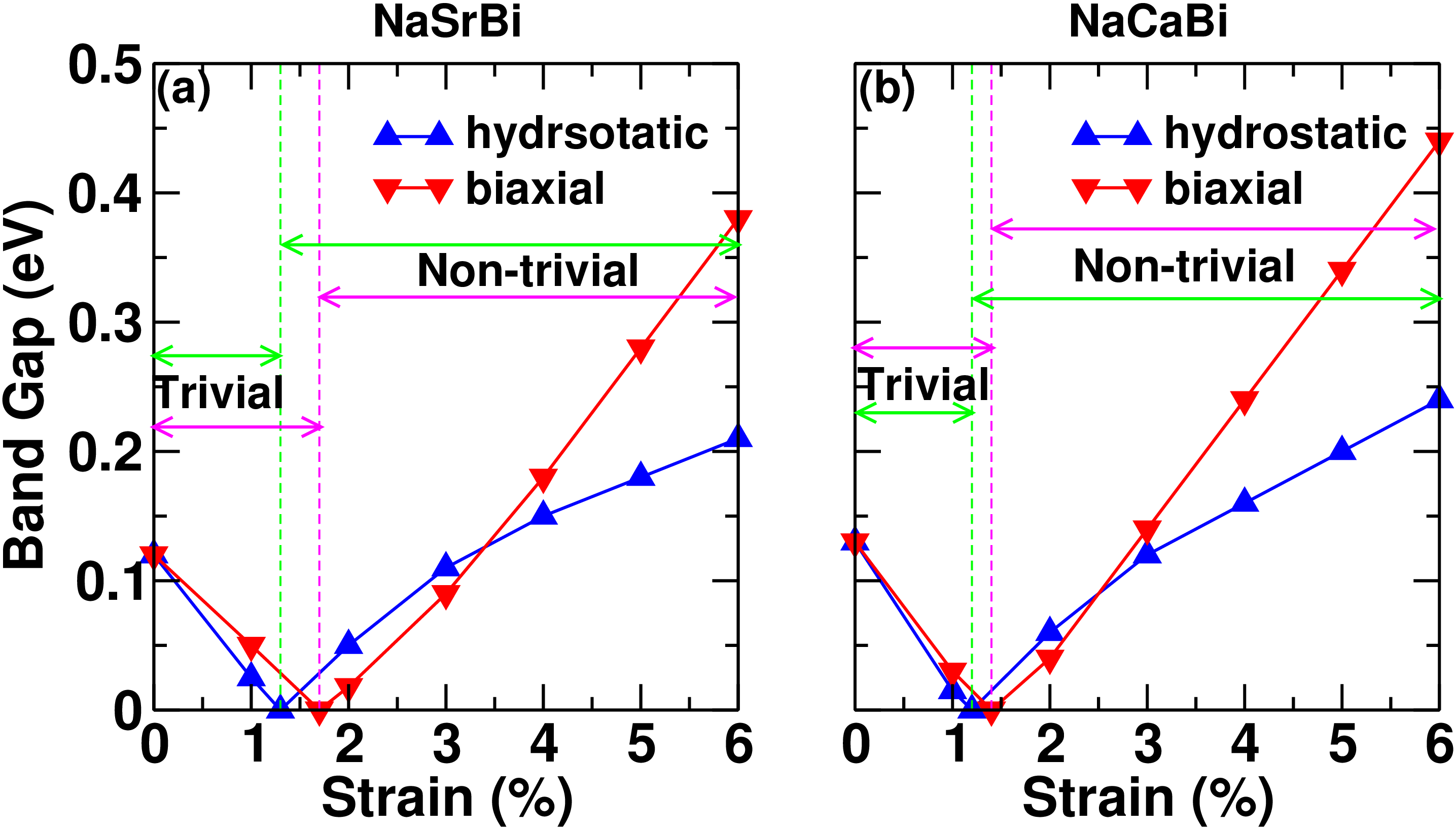}
\caption {Band gap vs. hydrostatic/bi-axial strain(\%) for (a) NaSrBi and (b) NaCaBi using HSE06 calculations. Trivial and non-trivial regions for both type of strains are marked by arrowheads.}
\label{fig3}
\end{figure}

{\par}{\it Hydrostatic Pressure}: We have performed electronic structure calculations on NaSrBi and NaCaBi systems under hydrostatic expansion. A trivial to non-trivial phase transition occurs at $\sim$ $-$2 GPa ($\sim$1$\%$ expansion in lattice parameter) and both the materials sustain non-trivial band order at higher expansion, as shown in Fig.~\ref{fig3}. Since  the calculated bulk modulus for NaSrBi and NaCaBi are 21 and 22 GPa respectively, it ensures that such non-trivial band ordering could be realized under low strain.  Interestingly, hydrostatic compression also gives non-trivial band ordering (band inversion between Bi-p and Ca/Sr d bands) in these systems. Our calculations show that p-d band inversion can be realized under a large hydrostatic compression (around $\sim\ 20$ GPa).  The detailed informations of compressive strain and the associated bands are given in SM\cite{supl} (Fig. S4). These pressure, however, are quite large and may not be easy to realize. Hence we consider only the hydrostatic expansion and investigated the non-trivial properties of both materials under 3$\%$ lattice expansion. Detailed bulk band structure and surface dispersions for both the systems with GGA and HSE06 level are shown in Sec. III of SM.\cite{supl}

{\par}{\it Bi-axial strain}: Next, we have investigated the electronic properties of these materials under biaxial strain (BAS). Experimentally, biaxial strain can be realized by substrate induced lattice mismatch. Accordingly, we have applied biaxial strain along [110] direction to observe the band evolution around the Fermi level. Figure ~\ref{fig3} shows the change in the band gap and trivial to non-trivial transform under biaxial strain for NaSrBi and NaCaBi. The trivial and non-trivial regions are mentioned in the plots using arrowheads. Above 1.6\% (1.4\%) biaxial strain, band inversion occurs in NaSrBi (NaCaBi), which sustains its non-trivial band ordering even at higher strain. Furthermore, we have simulated the bulk band structure for NaSrBi and NaCaBi at +3\% biaxial strain as shown in Fig.~\ref{fig2}. Fig.~\ref{fig2}(b,e) clearly shows band inversion between Sr/Ca $s$ and Bi $p_{z}$ orbitals at $\Gamma$ point. To further confirm the topological non-trivialness, we have calculated the topological $Z_2$ invariant. Owing to the inversion asymmetry in the crystal structure, the parity is not a good quantum number of the Bloch eigenstates. Therefore, parity counting method proposed by Fu and Kane is not applicable here. \cite{rd_1} As such we have adopted the method of Wannier charge center (WCC) evolution in half BZ to calculate the $Z_2$ invariant along the k$_2$ direction, as shown in Fig.~\ref{fig2}(c,f). It is clear from the figure that the WCC evolution lines cut the reference line one (odd) and zero (even) times in the k$_2$ = 0 and $\pi$ planes respectively, for both the systems. Thus the pressure-induced  band inversion exhibits a topological phase transition from a trivial insulator to TI.

\begin{figure}[b!]
\centering
\includegraphics[width=\linewidth]{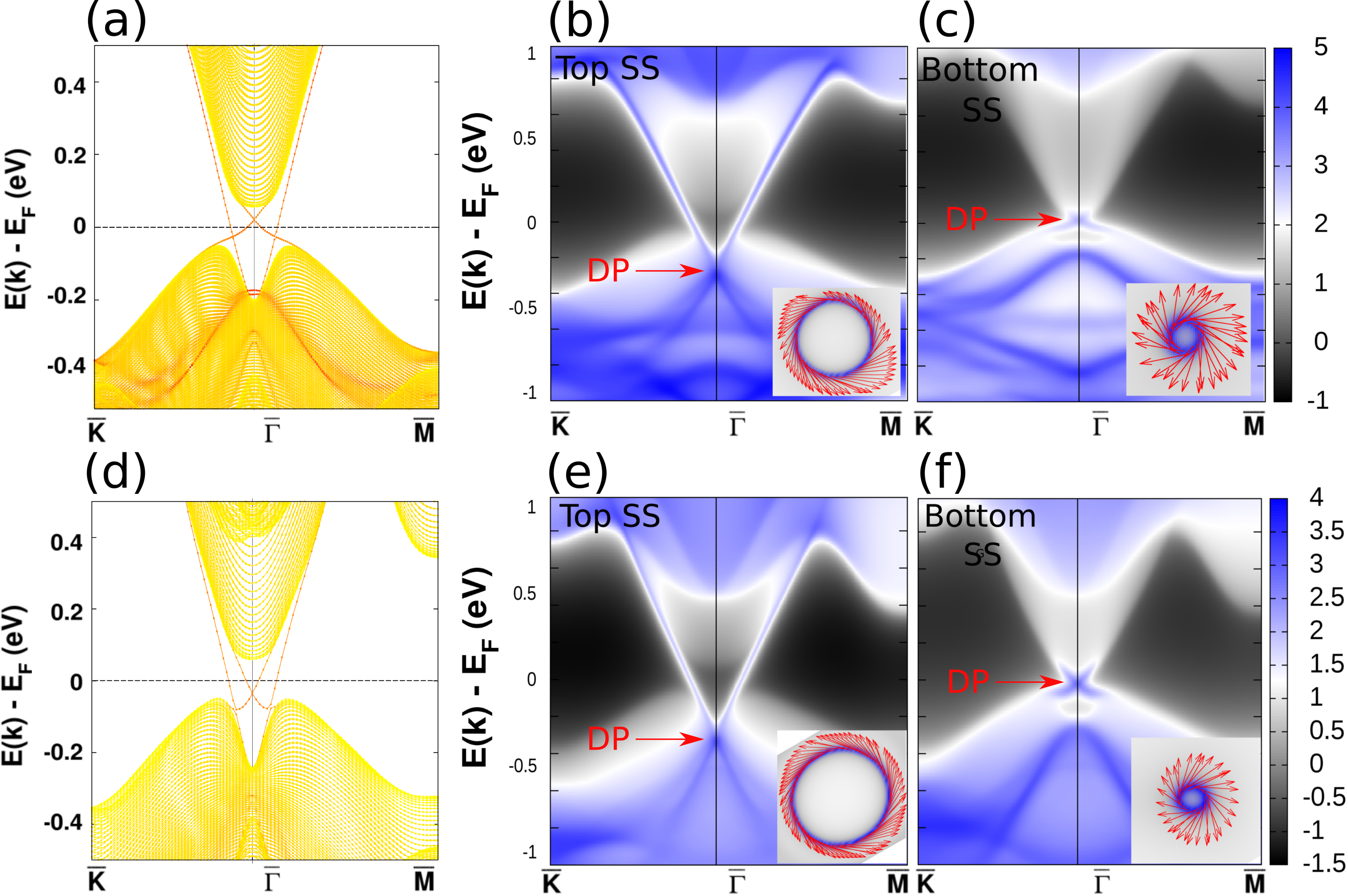}
\caption {(a) Surface dispersion, and surface density of states (SS) for (b) top, and (c) bottom surface layers of NaSrBi under +3\% BAS along [110]. (d-f) Respective plots for NaCaBi. The corresponding spin textures around the Fermi energy are shown in their respective insets.}
\label{fig4}
\end{figure}

{\par} To see the topological features, we have calculated surface spectra for NaSrBi and NaCaBi at +3\% BAS along [110]. The calculated bulk electronic structures using GGA and HSE06 show similar band ordering for both the systems (see Fig.~S5 of SM\cite{supl}). Hence it is reasonable to expect similar surface dispersion at GGA and HSE06 level of calculations. Therefore, we took the GGA functional to construct the MLWFs and then simulated the surface dispersions for TI phases of two compounds at +3\% BAS along [110]. The surface dispersion is shown in Fig.~\ref{fig4}. Since the slab calculation involves two surfaces, the corresponding surface bands and spectral intensity maps for both surfaces (top and bottom) are given.

{\par}In the slab model, the top surface is terminated by a X-Bi layer, while the bottom surface is truncated at X$^{'}$-Bi layer. The asymmetric surface truncation leads to different surface potentials which in turn results into two non-identical Dirac cones lying at different energy as shown in Fig.~\ref{fig4}. Another characteristic feature of topological surface state is the helical spin texture. To address this, we have projected the spin directions on the FS of the slab, which is located just above the DP and we find a spin momentum locking feature as shown in Fig.~\ref{fig4}. This again confirms the topological non-trivial behavior in both the systems. Similar to most other TI materials, the surface Dirac Cone of both the systems exhibits lefthanded spin texture for top surface states (TSS). The bottom surface states, however, exhibit righthanded spin texture for the Dirac Cone in both the materials.

{\par}{\it Doping}: Doping or alloying is a promising strategy for hydrostatic expansion/compression of lattice parameters. Therefore, we have doped K, Rb, and Cs at the Na site. Doping with bigger atoms leads to an expansion of lattice parameters, which in turn naturally causes a band inversion instead of a physical hydrostatic expansion imposed on the material. A detailed analysis of such findings, by doping K, Rb or Cs at Na sites in both NaSrBi and NaCaBi are discussed in SM\cite{supl} (see Fig. S6). 

{\it Nodal line semimetal:}

{\par} In a topological nodal line semi-metal, the bands cross each other due to band inversion and they form a closed loop instead of discrete points around the Fermi level. In  contrast to  WSs, which have an open arc like FS,\cite{in_16} NLSs are characterized  by the 1D closed ring (a line shape FS) and 2D topological drumhead surface state.\cite{in_17} The distinguishing characteristic of these drumhead surface state is that they are nearly dispersionless and therefore, have a large density of states near E$_F$. Such flat bands and large density of states could provide a potential play ground for the high temperature superconductivity, magnetism, and other related phenomenons.

\begin{figure}[t!]
\centering
\includegraphics[width=\linewidth]{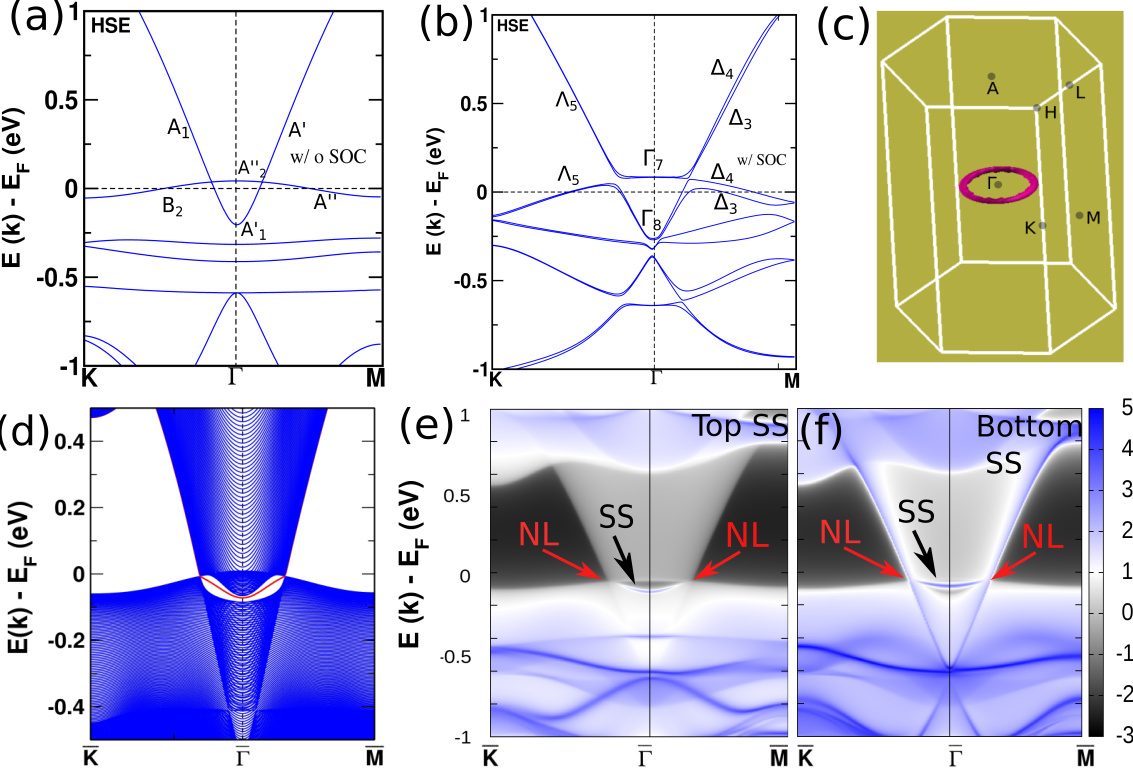}
\caption {(a-b) Bulk electronic structures (using HSE06) of RbCaBi without and with SOC. (c) Fermi surface plot for NLS phase of RbCaBi, and (d-f) surface dispersions and surface density of states (top and bottom) without SOC projected onto (001) surface.}
\label{fig5}
\end{figure}

{\par}Here we demonstrate that the materials NaSrBi (NaCaBi) can be transformed into a nodal line semi-metal by complete replacement of Na atom by Rb or Cs. Our detailed calculations predict that the class of systems XX$^{'}$Bi (X = Rb, Cs and X$^{'}$ = Ca, Sr) are NLS and show drumhead-like surface flat band. Bulk band structures for all these systems are shown in SM\cite{supl} (see Sec. V). Of these, we have chosen RbCaBi for detailed analysis here. Figure ~\ref{fig5}(a,c) shows the band structure of RbCaBi with an inverted band order and 1-D torus like bulk Fermi surface (where conduction and valence band crosses each other along a line) respectively in the absence of SOC. At the $\Gamma$ point, CBM and VBM have $A^{\prime\prime}_{2}$ and $A^{\prime}_{1}$ representation of $D_{3h}$. Along $\Gamma$-M, it becomes $A^{\prime\prime}$ and $A^{\prime}$ representation of $C_{s}$ where as it takes $B_{2}$ and $A_{1}$ representation of $C_{2v}$ along $\Gamma$-K,  as indicated in Fig.~\ref{fig5}(a). Therefore, lowest conduction band and highest valence band cross each other along the nodal line and protects from opening up a gap because the band repulsion is prohibited by the bands symmetry. Other systems also show similar nature of band structure, confirming the NLS behaviour (see  SM Fig.S7 \cite{supl}).
 
{\par}From the perspective of bulk boundary correspondence, topologically non-trivial drumhead-like surface states are expected to appear either inside or outside the projected nodal loop on the surface of NLS RbCaBi. In order to calculate the surface states, we have constructed tight binding Hamiltonian using the method of MLWFs and the surface states are projected onto (001) surface using the iterative scheme of Green's function technique. Interestingly, we found a nearly flat surface bands which is nestled between two bulk Dirac cones on the (001) surface, as shown in Fig.~\ref{fig5}(d-f).

{\par} Further, we take SOC effect into consideration and found that very small gaps are opened along the nodal line, and the material becomes a small gap topological insulator. In double group representation, CBM and VBM split into two  singly degenerate $\Gamma_{3}$ and $\Gamma_{4}$ bands along $\Gamma$-M direction as shown in Fig.~\ref{fig5}(b). Due to the opening of a small gap in both the directions, RbCaBi can be realized as a small gap strong topological insulator. Although it open up a small gap, the dispersion is still like NLS and expected to realise the topological flat surface bands  in experiment.


{\it Conclusion:}
In summary, using the first principles calculations, we have predicted topologically non-trivial phases including nodal line semi-metal states in a series of compounds belonging to the class of XX$^{'}$Bi (X=Na, K, Rb, Cs;\ X$^{'}$=Ca, Sr). We closely engineer the topology of the bands by applying hydrostatic compression/expansion, bi-axial strain and external doping, which in turn helps to achieve non-trivial band order.  Non-triviality is further confirmed by investigating Wannier charge center, surface dispersion and spin texture. NaSrBi and NaCaBi are found to be strong TI under hydrostatic and bi-axial strain. Doping or alloying is another efficient way to control the non-trivial order. Partial or complete replacement of Na by Rb, Cs or K in the compound NaX$^{'}$Bi (X$^{'}$=Sr, Ca) helps to intrigue the TI or the NLS phase. We have also studied 1-D bulk Fermi surface and the topological flat surface band properties of systems showing NLS behavior. Possibility of experimental synthesis is confirmed by presenting the chemical stability of all the compounds. We endorse a much higher predictability power of the present report due to the use of HSE06 functionals compared to most of the similar previous reports based on GGA functional. Such accurate ab-initio predictions serve as a guiding path for the discovery of new novel materials.

{\it Acknowledgements:}
This work is financially supported by DST SERB (EMR/2015/002057), CSIR, India. We thank IIT Indore for the lab and computing facilities. C.M, C.K.B and S.K thank MHRD for research fellowship. AA acknowledges National Center for Photovoltaic Research and Education (NCPRE), IIT Bombay for possible funding to support this research.

\end{document}